\documentclass[prl,twocolumn,superscriptaddress,showpacs,times,10pt]{revtex4-1}
\usepackage{graphicx,amsmath,amssymb,bm}
\usepackage{color}
\usepackage{array}
\usepackage{rotating,pifont}
\usepackage{simplewick}

\DeclareMathSymbol{\NS}{\mathord}{AMSb}{"4E}

\newcommand{\hw}{\ensuremath{\hbar\omega}}
\newcommand{\lambdaSRG}{\ensuremath{\lambda_{\text{SRG}}}}

\newcommand{\fmi}{\ensuremath{\,\text{fm}^{-1}}}

\newcommand{\keV}{\ensuremath{\,\text{keV}}}
\newcommand{\MeV}{\ensuremath{\,\text{MeV}}}

\definecolor{FGViolet}{rgb}{0.61,0.32,0.61}
\definecolor{FGDarkBlue}{rgb}{0,0,0.6}
\definecolor{FGBlue}{rgb}{0,0,0.8}
\definecolor{FGLightBlue}{rgb}{0.2, 0.6, 0.8}
\definecolor{FGGreen}{rgb}{0.2,0.7,0.2}
\definecolor{FGLightGreen}{rgb}{0.4,1,0.4}
\definecolor{FGYellow}{rgb}{1,0.95,0}
\definecolor{FGOrange}{rgb}{0.95,0.5,0.1}
\definecolor{FGRed}{rgb}{0.8,0,0}
\definecolor{FGWhite}{rgb}{1,1,1}
\definecolor{FGLightGray}{rgb}{0.8,0.8,0.8}
\definecolor{FGGray}{rgb}{0.5,0.5,0.5}
\definecolor{FGDarkGray}{rgb}{0.3,0.3,0.3}
\definecolor{FGBlack}{rgb}{0,0,0}

\begin{document}

\title{Ground and excited states of doubly open-shell nuclei \\
from ab initio valence-space Hamiltonians}

\author{S.\ R.\ Stroberg}
\email[E-mail:~]{sstroberg@triumf.ca}
\affiliation{TRIUMF, 4004 Wesbrook Mall, Vancouver, British Columbia, 
V6T 2A3 Canada}

\author{H.\ Hergert}
\email[E-mail:~]{hergert@nscl.msu.edu}
\affiliation{National Superconducting Cyclotron Laboratory and 
Department of Physics and Astronomy, Michigan State University,
East Lansing, MI 48824, USA}

\author{J.\ D.\ Holt}
\email[E-mail:~]{jholt@triumf.ca}
\affiliation{TRIUMF, 4004 Wesbrook Mall, Vancouver, British Columbia, 
V6T 2A3 Canada}

\author{S.\ K.\ Bogner}
\email[E-mail:~]{bogner@nscl.msu.edu}
\affiliation{National Superconducting Cyclotron Laboratory and 
Department of Physics and Astronomy, Michigan State University, 
East Lansing, MI 48824, USA}

\author{A.\ Schwenk}
\email[E-mail:~]{schwenk@physik.tu-darmstadt.de}
\affiliation{Institut f\"ur Kernphysik, Technische Universit\"at Darmstadt, 
64289 Darmstadt, Germany}
\affiliation{ExtreMe Matter Institute EMMI, GSI Helmholtzzentrum f\"ur
Schwerionenforschung GmbH, 64291 Darmstadt, Germany}

\begin{abstract}

We present ab initio predictions for ground and excited states of doubly 
open-shell fluorine and neon isotopes based on chiral two- and 
three-nucleon interactions. We use the in-medium similarity 
renormalization group, to derive mass-dependent $sd$ valence-space 
Hamiltonians. The experimental ground-state energies are reproduced 
through neutron number $N=14$, beyond which a new targeted 
normal-ordering procedure improves agreement with data and 
large-space multi-reference calculations. For spectroscopy, we focus on 
neutron-rich $^{23-26}$F and $^{24-26}$Ne isotopes near $N=14,16$ 
magic numbers. In all cases we find agreement with experiment and
established phenomenology.  Moreover, yrast states are well described in 
$^{20}$Ne and $^{24}$Mg, providing a path towards an ab initio 
description of deformation in the medium-mass region.

\end{abstract}


\maketitle

%\paragraph{Introduction.}

With hundreds of undiscovered nuclei to be created and studied at 
rare-isotope beam facilities, the development of an ab initio picture of 
exotic nuclei is a central goal of modern nuclear theory. Three-nucleon 
(3N) forces are a key input to understand and predict the structure of 
medium-mass nuclei, from the neutron dripline in oxygen to the evolution 
of magic numbers in oxygen and calcium
\cite{Hebe15ARNPS,Otsu10Ox,Hage12Ox3N,Cipo13Ox,Herg13IMSRG,Herg13MR,Holt12Ca,Hage12Ca3N,Gall12Ca,Wien13Nat,Holt13PR}.  
In addition, advances in large-space many-body methods have extended 
the scope of ab initio theory to open-shell calcium and nickel isotopes, and 
beyond \cite{Soma14GGF,Bind14CCheavy,Herg14MR}.  While 
ground-state properties of even-even isotopes are captured with these 
methods, excited states and/or odd-mass systems away from closed 
shells are more challenging. Furthermore, doubly open-shell nuclei may 
exhibit deformation, which is challenging to capture in large-space ab initio 
methods built on spherical reference states 
\cite{Sign15BCC,Duge15SRCC}.  

These difficulties can be addressed straightforwardly within the framework 
of the nuclear shell model
\cite{Brow01PPNP,Caur05RMP,Otsu13PS}, where an effective 
valence-space Hamiltonian is constructed for particles occupying a small 
singe-particle space above some closed-shell configuration. Exact 
diagonalization then accesses all nuclei and their structure properties in a 
given region and naturally captures deformation \cite{Elli58SMdef}. While 
the shell model approach is traditionally phenomenological, valence-space 
Hamiltonians obtained with many-body perturbation theory (MBPT) 
\cite{Hjor95MBPT} including 3N forces describe separation energies and 
first-excited $2^+$ energies in the $sd$ shell above $^{16}$O 
\cite{Gall14IMME,Simo15unc}. However, order-by-order convergence of 
is difficult to verify, especially for $T = 0$ components, and a successful 
description of exotic nuclei requires the use of extended valence spaces 
\cite{Holt13Ox,Holt14Ca,Tsun14EKK,Dong14KK}. All-order diagrammatic 
extensions provide further insights \cite{Holt05CP} but exhibit dependence 
on the harmonic-oscillator spacing $\hw$ and have not been benchmarked 
with 3N forces. Recently, nonperturbative methods have been developed
\cite{Tsuk12SM,Bogn14SM,Jans14SM,Lise08NCSMSM,Dikm15NCSMSM}, 
which provide a promising path toward an ab initio description of nuclei 
between semi-magic isotopic chains, but have not been applied 
systematically beyond oxygen.

\begin{figure*}[t]
\includegraphics[scale=.7,clip=]{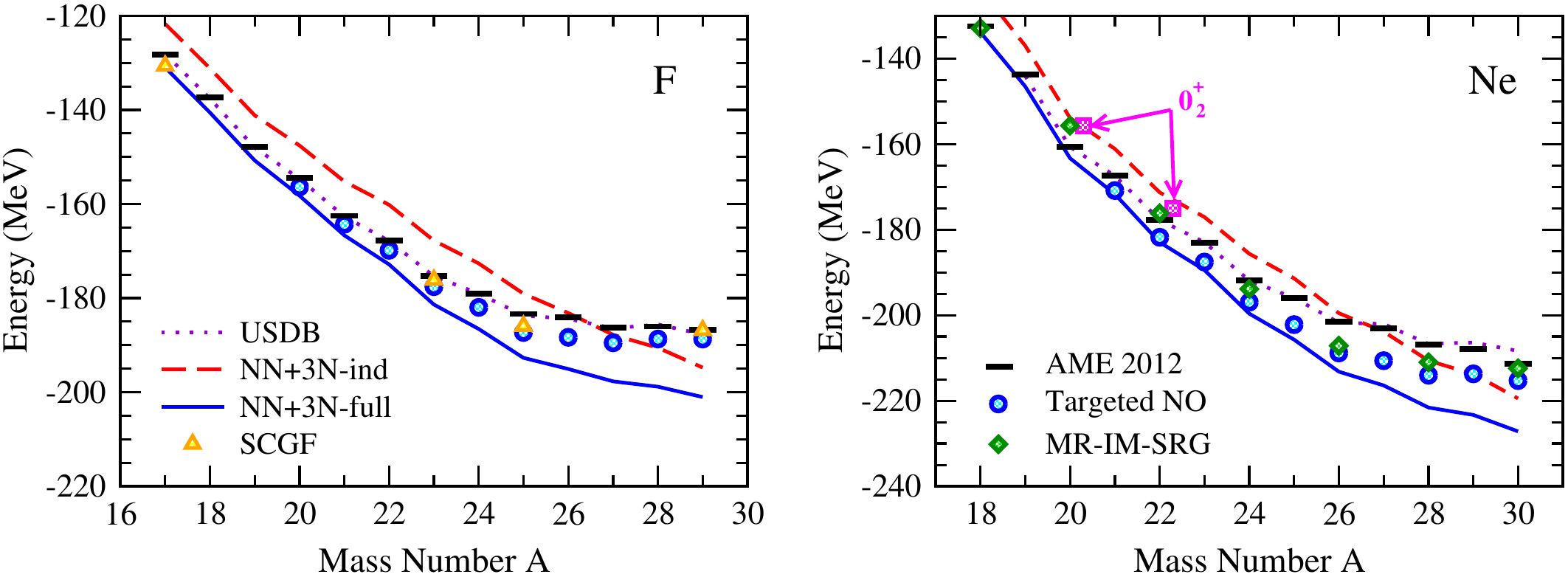}
\caption{Ground-state energies of fluorine and neon isotopes from the 
$A$-dependent IM-SRG valence-space Hamiltonian with 
$\lambdaSRG = 1.88 \fmi$ and $\hw=24\MeV$ compared with the 2012 
Atomic Mass Evaluation (AME2012) \cite{Wang12AME12} and the 
phenomenological USDB interaction \cite{Brow06USD}.  Blue circles 
indicate results obtained with the new targeted normal ordering (Targeted NO) scheme, yellow triangles self-consistent Green's function (SCGF) \cite{Cipo13Ox}, and green diamonds 
indicate ground-state energies calculated with the multi reference (MR-IM-SRG).
\label{gs}}
\end{figure*}

In this article we present ab initio predictions for ground and excited states 
in doubly open-shell nuclei using valence-space Hamiltonians derived 
from the in-medium similarity renormalization group (IM-SRG). Focusing 
on fluorine and neon isotopes within the $sd$-shell, we find that including 
chiral 3N forces leads to a good agreement with experimental data and 
state-of-the-art phenomenology \cite{Brow06USD}.  We also introduce a 
novel targeted normal-ordering procedure, which further improves 
ground-state energies in comparison to experiment and large-space 
multi-reference IM-SRG calculations performed directly in the target 
nucleus. Finally we demonstrate that nuclear deformation in medium-mass 
nuclei emerges ab initio by studying yrast states in $^{20}$Ne and 
$^{24}$Mg and comparing with spherical ground states obtained with
multi-reference IM-SRG \cite{Herg13MR}.

%\paragraph{In-Medium SRG.}

In the IM-SRG, we start from an $A$-body Hamiltonian that is normal 
ordered with respect to a finite-$A$ reference, e.g., a Hartree-Fock ground 
state, and apply a continuous unitary transformation $U(s)$ to drive the 
Hamiltonian to band- or block-diagonal form.  In practice, this is  
accomplished by solving the flow equation
\begin{equation}
\label{eq:srg}
\frac{dH(s)}{ds} =[\eta(s),H(s)]\,, 
\end{equation}
where $U(s)$ is defined implicitly through the anti-Hermitian generator 
$\eta(s) \equiv [dU(s)/ds]\,U^{\dagger}(s)$.  With a suitable choice of 
$\eta(s)$, the off-diagonal part of the Hamiltonian, $H^{\rm{od}}(s)$, is 
driven to zero as $s \rightarrow \infty$. The freedom in defining 
$H^{\rm{od}}(s)$ allows us to tailor the decoupling to the problem of 
interest, e.g., the core \cite{Tsuk11IMSRG,Herg13IMSRG} or the 
core and a valence-space Hamiltonian \cite{Tsuk12SM,Bogn14SM}. 
Within the IM-SRG(2) approximation, Eq.~\eqref{eq:srg} is truncated to 
normal-ordered two-body operators.  In the present work, we use a version 
of White's generator which is less susceptible to the effects of small energy 
denominators than the one we used in earlier work
\cite{Tsuk12SM,Bogn14SM}. Denoting generic energy denominators by 
$\Delta$, $\eta = 1/2 \tan^{-1}(2H^{\mathrm{od}}/\Delta)$ 
\cite{Whit02SRG}. We also apply the newly developed Magnus formulation 
\cite{Morr15Magnus} to decouple valence-space Hamiltonians, where the 
unitary transformation $U(s)$ is explicitly calculated, making the 
calculation of general effective operators for observables such as radii or 
electroweak transitions tractable. Results calculated within both 
frameworks agree at the $10 \keV$ level for both core and valence-space 
decoupling. 

%\paragraph{Implementation.}

To implement the IM-SRG, we start from the 
$\Lambda_{\rm{NN}}=500\MeV$ chiral N$^3$LO NN interaction of 
Refs.~\cite{Ente03EMN3LO,Mach11PR} and evolve with the free-space 
SRG \cite{Bogn07SRG,Bogn09PPNP} to low-momentum resolution 
scales, $\lambdaSRG=1.88-2.11\fmi$. The NN+3N-induced 
(NN+3N-ind) Hamiltonians includes 3N forces induced by the evolution 
and correspond to the original NN interaction, up to neglected induced 
four- and higher-body forces \cite{Bogn09PPNP,Jurg09SRG3N}. The 
NN+3N-full Hamiltonians include an initial local 
$\Lambda_{\rm 3N} = 400 \MeV$ chiral N$^2$LO 3N interaction 
\cite{Navr07local3N}, consistently evolved to $\lambdaSRG$. This value 
of $\Lambda_{\rm 3N}$ minimizes the effects of induced 4N interactions 
in the region of oxygen
\cite{Roth12NCSMCC3N,Roth14SRG3N,Bogn14SM}. Calculations in 
oxygen isotopes with $\Lambda_{\rm 3N} = 500 \MeV$ displayed a 
pronounced sensitivity to $\lambdaSRG$ \cite{Bogn14SM}, making it 
difficult to disentangle uncertainties originating from neglected induced 
forces and the initial Hamiltonian. To obtain the final input Hamiltonian, we 
add the $A$-dependent intrinsic kinetic energy. Here, we choose $A$ to
be the mass of the target nucleus, for which we wish to approximate
an exact no-core diagonalization. An $A$-independent prescription 
introduces an error that grows with the number of valence nucleons
\cite{Stro15Adep}.

We then solve the Hartree-Fock equations to obtain the core reference
state. We normal order the Hamiltonian with respect to the Hartree-Fock 
reference state and discard residual three-body forces 
\cite{Hage07CC3N}. The normal-ordered 0-, 1-, and 2-body parts are 
taken as initial values in the IM-SRG decoupling within a single-particle 
basis $e=2n+l \le e_{\mathrm{max}}=14$, with an additional cut 
$e_1+e_2+e_3 \le E_{\mathrm{3max}}=14$ for 3N forces
\cite{Roth14SRG3N}.  

The IM-SRG is used to decouple the core and valence space from 
excitations, and the core energy, valence-space single-particle energies, 
and two-body matrix elements are taken from the evolved 
$s \rightarrow \infty$ Hamiltonian \cite{Tsuk12SM,Bogn14SM}.  We work 
within the standard $sd$ shell consisting of the proton and neutron 
$d_{5/2}$, $d_{3/2}$, and $s_{1/2}$ orbits above the $^{16}$O 
core. We diagonalize the $A$-dependent valence-space Hamiltonian to 
obtain ground-state energies and natural-parity spectra using the NushellX 
and Oslo shell model codes \cite{Brow14Nushx,OsloCode}. Since it is well 
known that the NN+3N-full initial Hamiltonian used in these calculations 
produces systematic overbinding and too-small radii in calcium 
\cite{Soma14GGF,Bind14CCheavy,Herg14MR}, we limit our discussion to 
isotopic chains in the lower $sd$ shell, in particular fluorine and neon, 
which serve to test proton-proton, neutron-neutron, and proton-neutron 
valence-space matrix elements. With increasing valence particle number, 
ab initio valence-space Hamiltonians must also systematically account for 
3N forces within the valence space, an issue we address when discussing 
our targeted normal ordering approach.  

\begin{figure*}
\begin{center}
\minipage{0.25\textwidth}
\includegraphics[width=\linewidth,clip=]{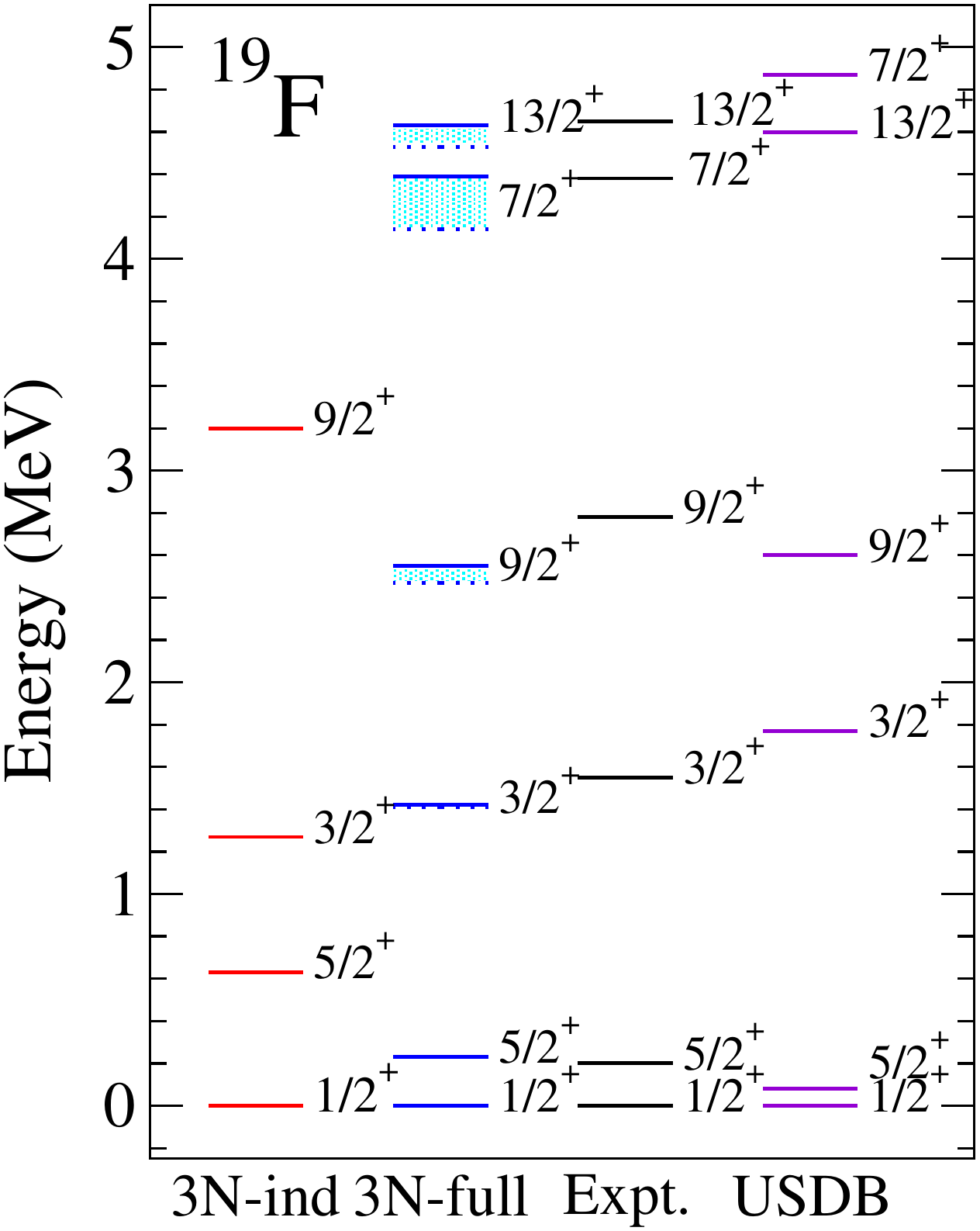}
\endminipage\hfill
\minipage{0.225\textwidth}
\includegraphics[width=\linewidth,clip=]{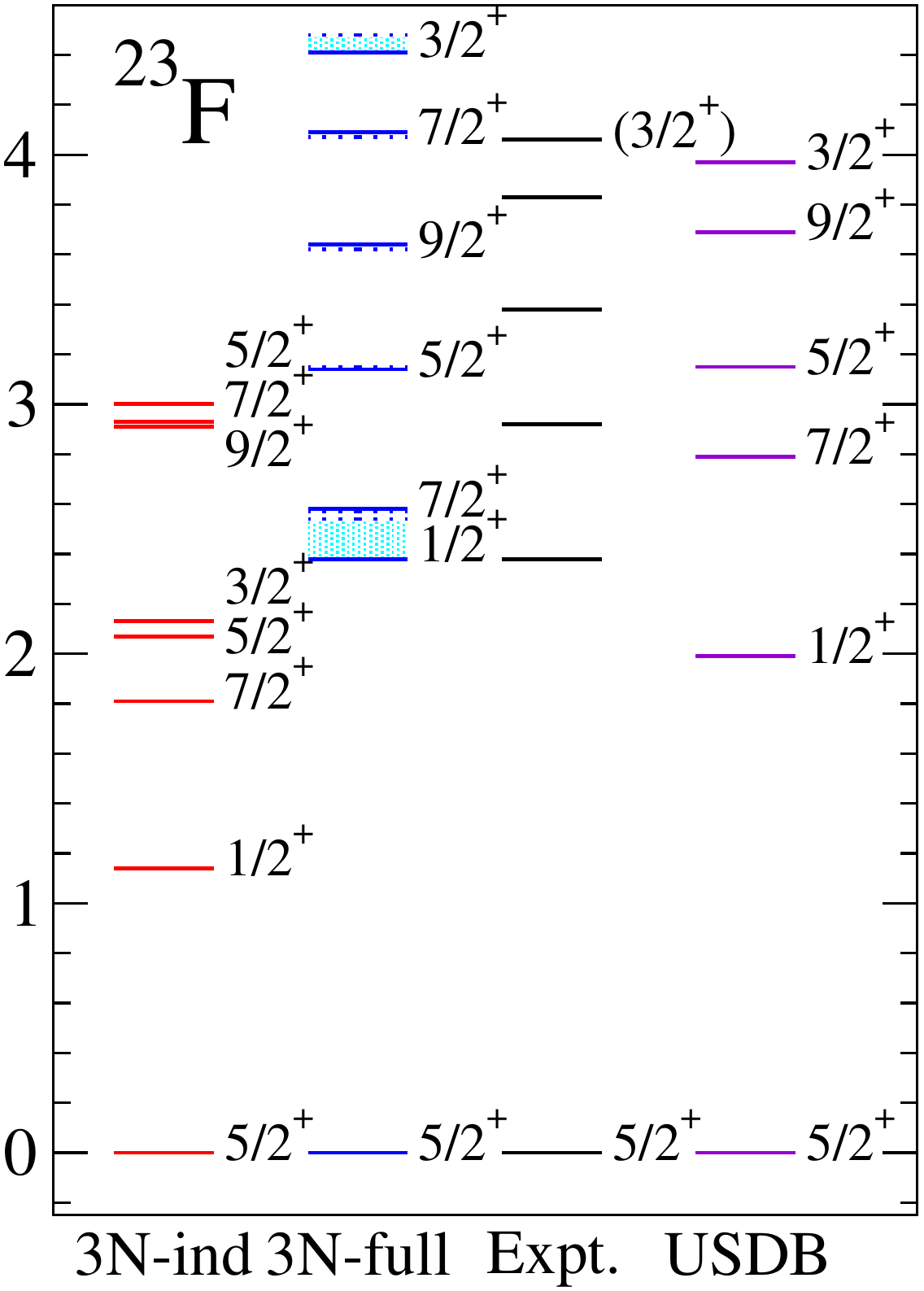}
\endminipage\hfill
\minipage{0.225\textwidth}
\includegraphics[width=\linewidth,clip=]{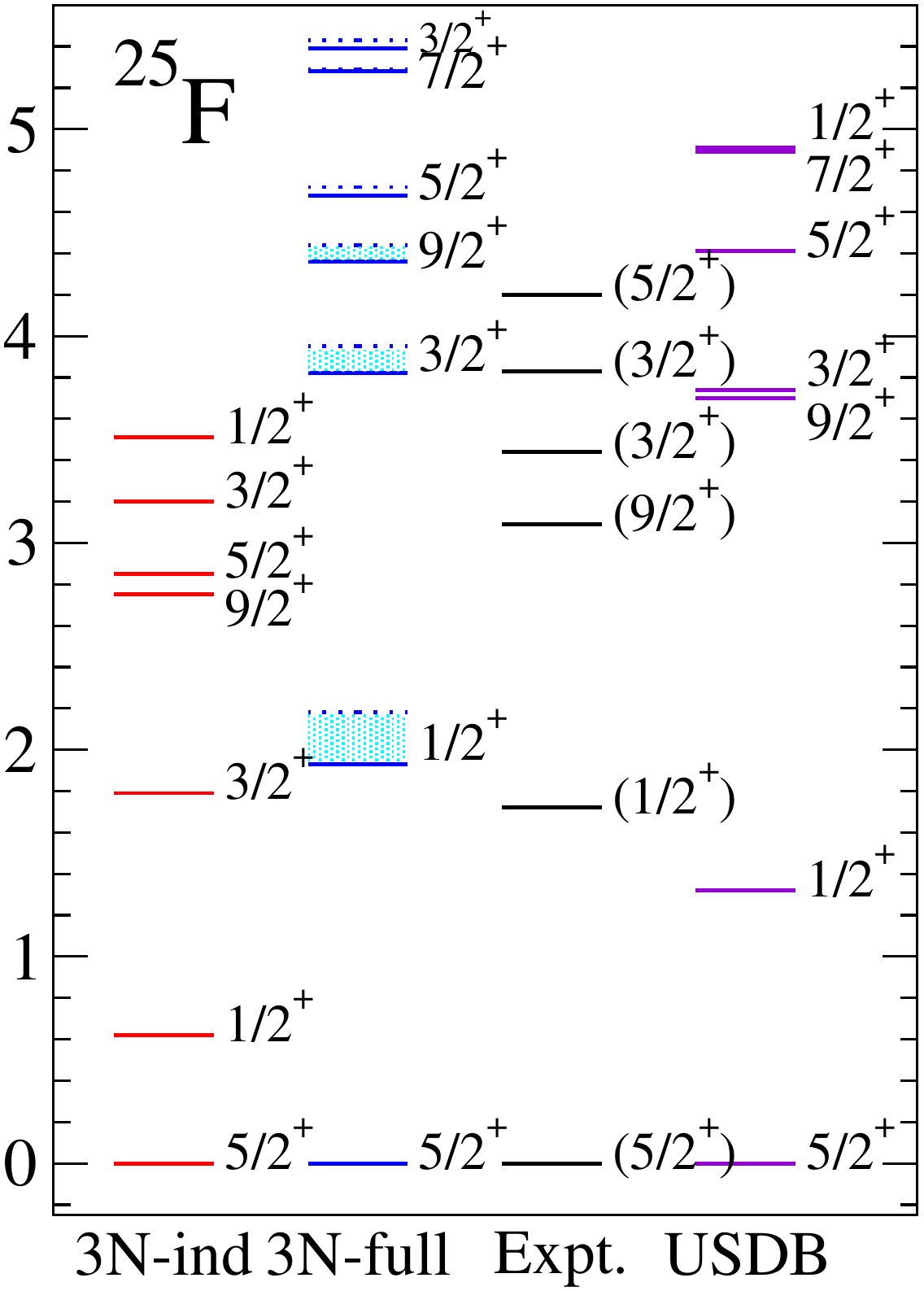}
\endminipage\hfill
\minipage{0.225\textwidth}
\includegraphics[width=\linewidth,clip=]{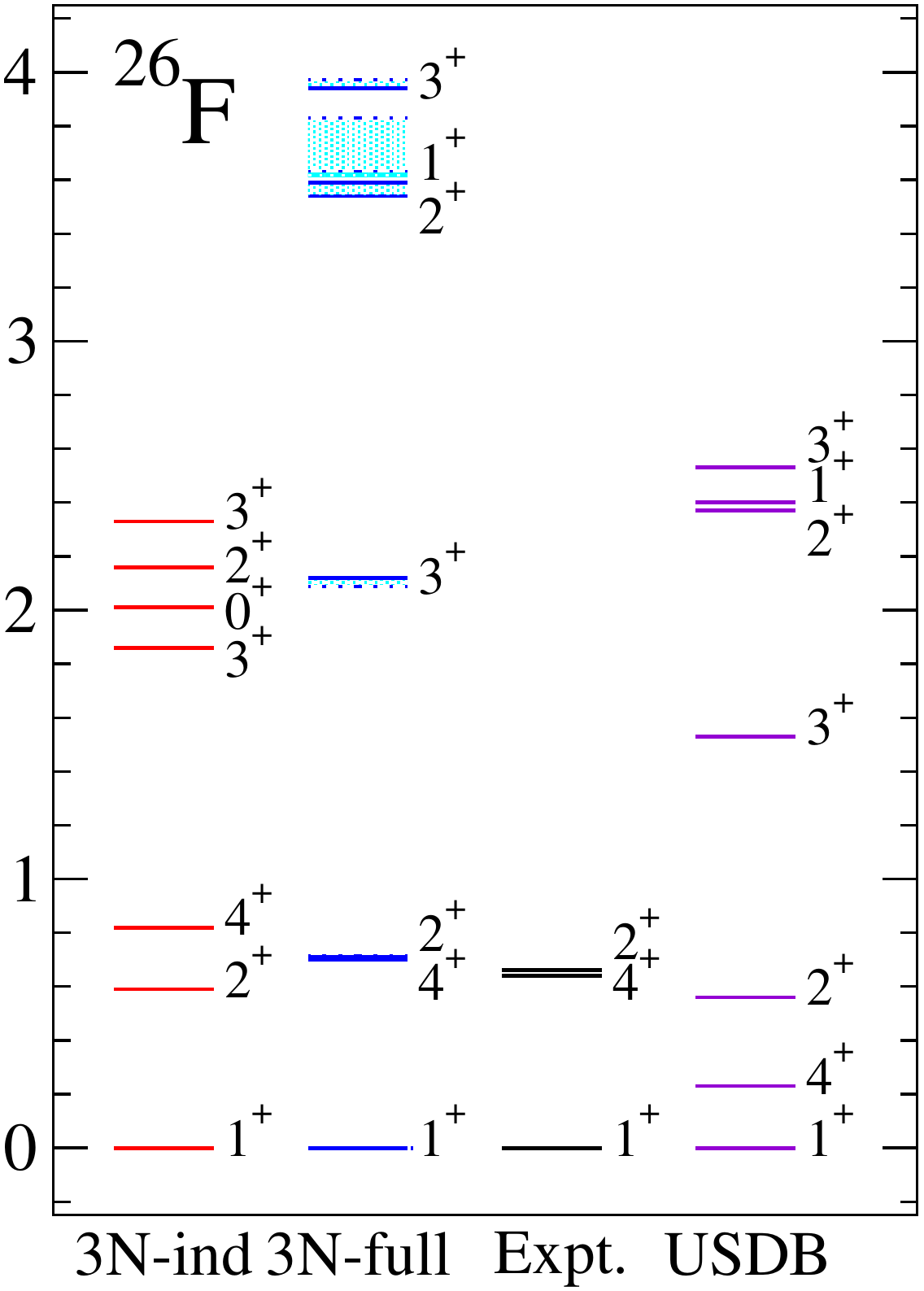}
\endminipage
\end{center}
\caption{Excited-state spectra for $^{19,23,25,26}$F from
IM-SRG Hamiltonians based on NN+3N-ind and NN+3N-full Hamiltonians 
for $\Lambda_{\mathrm{3N}}=400\MeV$ with $\hw=20\MeV$ (dotted) and 
$\hw=24\MeV$ (solid), compared with experiment \cite{nndc14ENSDF} 
and results from the phenomenological USDB interaction 
\cite{Brow06USD}.}
\label{Spectra:F}
\end{figure*}

%\paragraph{Results.}

We first consider ground-state energies in fluorine and neon isotopes, 
which have been explored with self-consistent Green's function
calculations for particular isotopes \cite{Cipo13Ox,Cipo14Ox} and 
valence-space Hamiltonians from MBPT 
\cite{Hebe15ARNPS,Simo15unc}. IM-SRG results through $N=20$ are 
shown in Fig.~\ref{gs}, compared with experiment and phenomenological 
USDB predictions \cite{Brow06USD}. Since core properties are 
calculated consistently in our IM-SRG framework, we quote absolute 
ground-state energies in all calculations, but normalize USDB results to 
the experimental ground state of $^{16}$O.  We first observe that 
NN+3N-ind Hamiltonians exhibit incorrect trends throughout both isotopic 
chains, reminiscent of the incorrect dripline predictions in oxygen isotopes 
\cite{Hebe15ARNPS,Otsu10Ox,Bogn14SM}.  With NN+3N-full 
Hamiltonians, the agreement is improved, including the flattening of 
energies in the neutron-rich region.  We note a very minor $\hw$ 
dependence of ground-state energies for $\hw\!=\!20\!-\!24\MeV$, not 
shown in Fig.~\ref{gs}.  The largest deviations are $600\keV$ in $^{29}$F 
and $1.3 \MeV$ in $^{30}$Ne, a 0.5\% effect, indicating good convergence 
with respect to the model space truncation.  

It is apparent, however, that near $N=14$, NN+3N-full results become 
overbound with respect to experiment, similar to oxygen isotopes 
\cite{Bogn14SM}. We also plot in Fig.~\ref{gs} multi-reference IM-SRG 
calculations of ground-state energies in even neon isotopes based on the 
same initial Hamiltonian, which display an improved agreement with 
experiment outside of $^{20,22}$Ne. One obvious difference between the 
valence-space and multi-reference formulations is that the latter is carried 
out in the target nucleus. In the valence-space calculations, the 
Hamiltonian is instead normal ordered with respect to the $^{16}$O core, 
which neglects 3N forces between valence nucleons. This approximation 
works well for few valence nucleons, but residual 3N effects scale as 
$A_v/A_c$ \cite{Frim113Nres} for normal Fermi systems, and therefore 
cannot be neglected as the number of valence nucleons increases 
\cite{Caes1326O,Holt14Ca}. 

To mitigate this effect, we introduce a targeted normal ordering approach 
in which the normal ordering is first performed with respect to the nearest 
closed shell rather than the $^{16}$O core. We then apply the IM-SRG to 
decouple the $^{16}$O core and $sd$ valence space. Finally, we 
re-normal order with respect to $^{16}$O to perform a full $sd$-shell 
diagonalization. The results of this procedure are shown in both figures, 
which provides $12\MeV$ additional repulsion at $N=20$ and improves 
agreement with experiment.  More importantly, there are only modest 
differences between the shell model results and large-space self-
consistent Green's function and multi-reference IM-SRG calculations in 
fluorine and neon, respectively.  Furthermore the impact on spectra is 
generally minor for both isotopic chains.  

\begin{figure*}
\begin{center}
\minipage{0.25\textwidth}
\includegraphics[width=\linewidth,clip=]{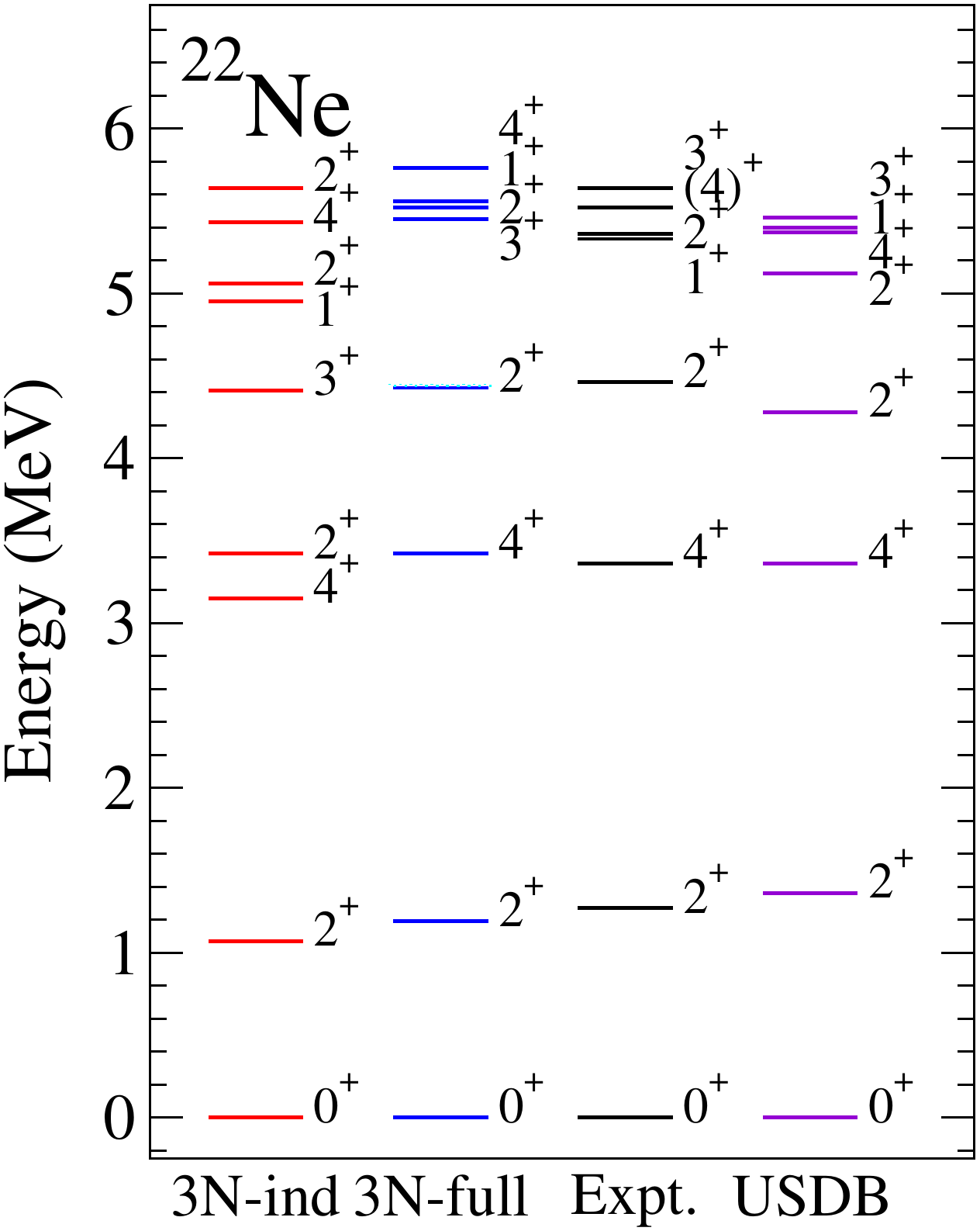}
\endminipage\hfill
\minipage{0.225\textwidth}
\includegraphics[width=\linewidth,clip=]{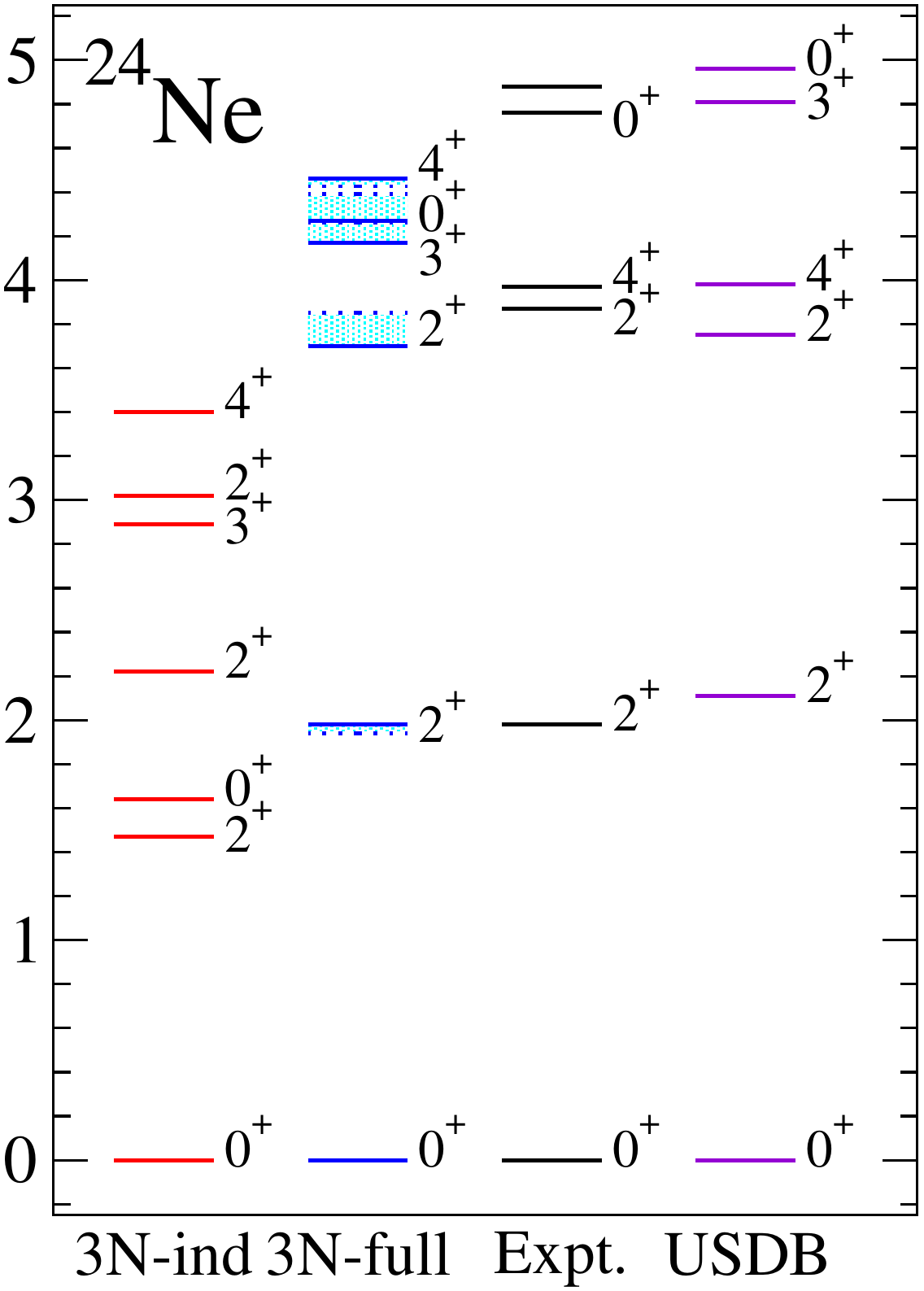}
\endminipage\hfill
\minipage{0.225\textwidth}
\includegraphics[width=\linewidth,clip=]{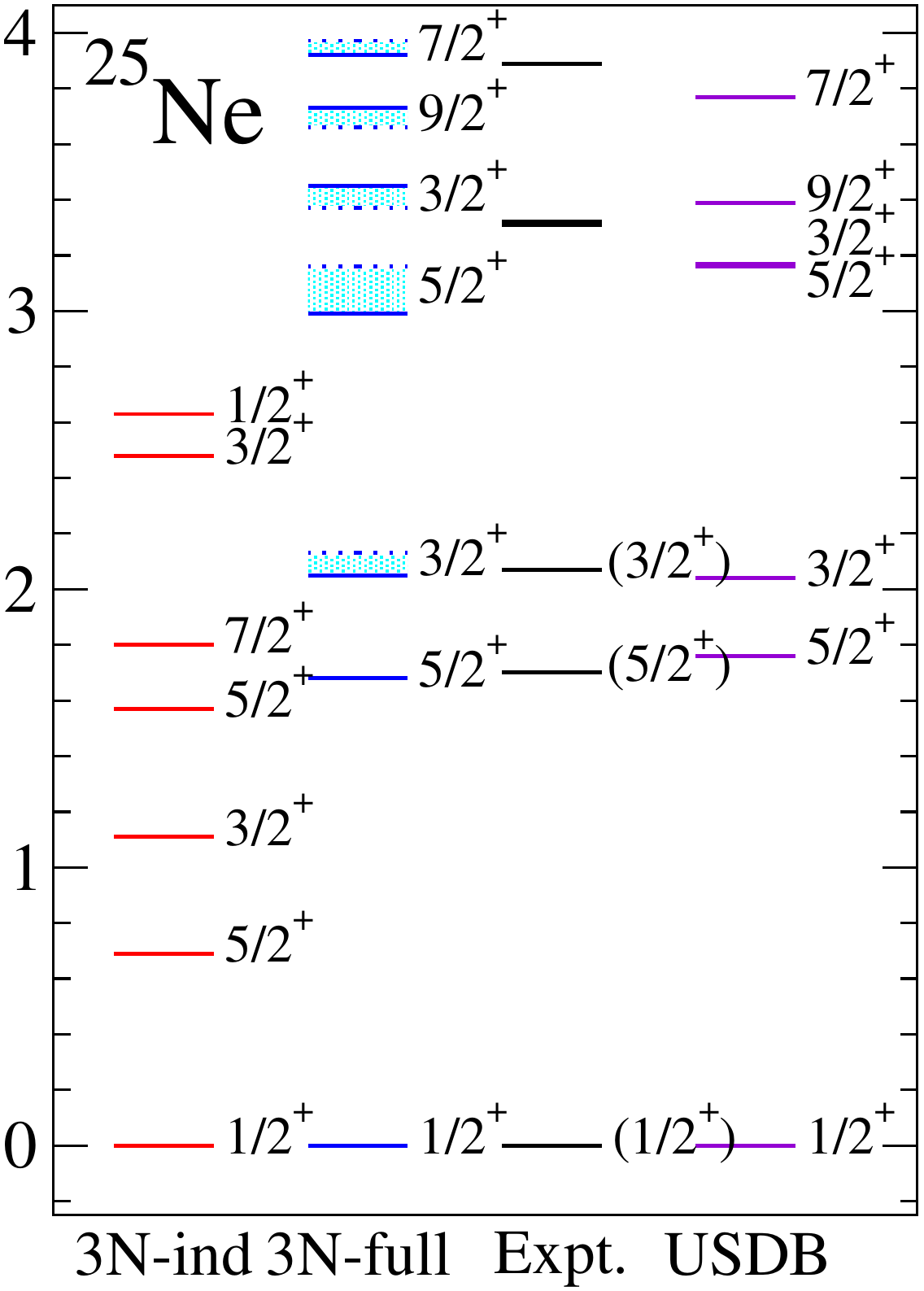}
\endminipage\hfill
\minipage{0.225\textwidth}
\includegraphics[width=\linewidth,clip=]{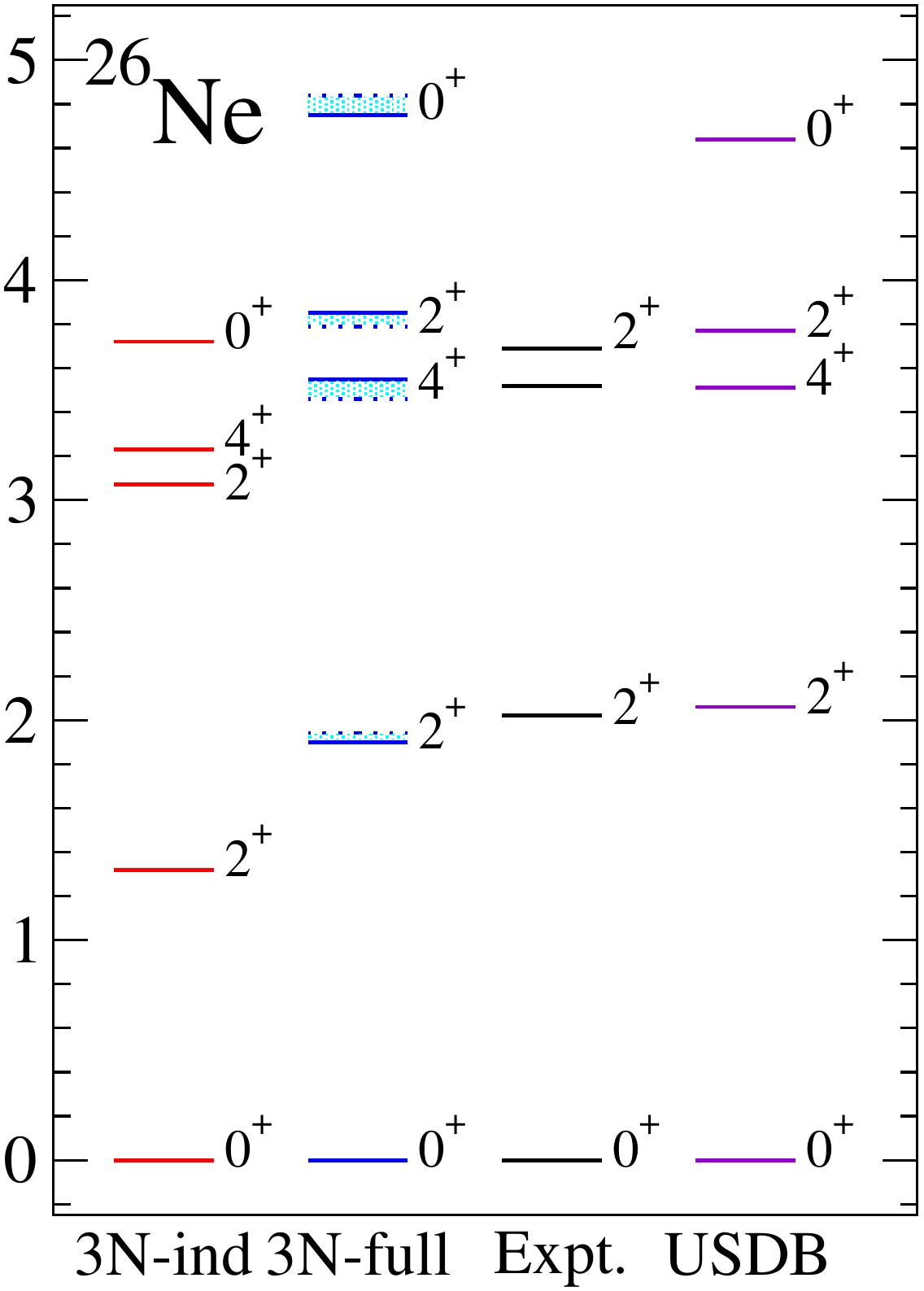}
\endminipage
\end{center}
\caption{Excited-state spectra of $^{19,24,25,26}$Ne, as in Fig.~
\ref{Spectra:F}.}
\label{Spectra:Ne}
\end{figure*}

%\subsection{Spectra}

For spectroscopy in the fluorine and neon isotopes, we highlight the 
$N=14,16$ region towards the experimental limits, in addition to one 
example at stability, though for completeness, we show spectra for all F, 
Ne, Na, and Mg isotopes within the $sd$ shell as supplementary material 
\cite{SupMat}, and interaction files are available online \cite{IntFiles}.  
The only ab initio predictions in fluorine are large-scale coupled-cluster 
calculations in $^{26}$F using a phenomenological 3N force 
\cite{Lepa1326F} and $^{22,24}$F using optimized chiral interactions at 
order N$^2$LO \cite{Ekst14GT2bc}. In both cases, spectra are 
reasonable, but the density and ordering of states can deviate from 
experiment \cite{Ekst14GT2bc,Hebe15ARNPS}. IM-SRG calculations in 
$^{24}$F succeeded in predicting properties of newly measured states 
\cite{Cace1524F}. There are no ab initio predictions for spectra in neon 
except for the first excited $2^+$ energies in even isotopes from MBPT 
shell model based on 3N forces \cite{Simo15unc}. Finally, we denote the 
$\hw$ dependence of spectra with shaded bands in NN+3N-full results. 
While often at the $100\keV$ level or less, in a few cases it approaches 
$400\keV$. 

In Fig.~\ref{Spectra:F} we show the calculated spectra of
$^{19,23,25,26}$F.  We first observe that in all cases, NN+3N-ind forces 
give too-compressed spectra with an incorrect ordering of levels, 
even in the stable $^{19}$F. With initial 3N forces, the spectra are clearly 
improved. The spectrum of $^{19}$F agrees very well with experiment, 
even giving the correct $7/2^+ \!- \! 13/2^+$ ordering not reproduced by 
USDB. For the neutron-rich isotopes, experimental data are fewer.
Nonetheless the spacing of the mostly unidentified levels in $^{23}$F are 
reproduced, and spin-parity assignments agree with USDB below 
$4 \MeV$.  In $^{25}$F neither IM-SRG nor USDB fully predict the 
experimental spectrum and, despite similar spacings, do not agree on the 
ordering of states.  Finally in $^{26}$F only the lowest excited states 
are known and are well reproduced by IM-SRG.  The ordering of 
higher-lying excited states agrees well with USDB, but the increased 
energy is likely due to a lack of continuum effects, which are implicitly 
included in the phenomenology.  Additional experimental spin/parity 
assignments are needed to conclusively test our predictions.

In Fig.~\ref{Spectra:Ne} we show calculations for the stable $^{22}$Ne
and exotic $^{24-26}$Ne nuclei. Experimental data are limited, but in all 
cases, spectra without initial 3N forces are too compressed with respect to 
experiment, particularly $^{25}$Ne. With initial 3N forces, the spectra are 
improved throughout the chain. For example in $^{25,26}$Ne the ordering 
of states is in complete agreement with USDB, strongly suggesting the 
unidentified excited state in $^{26}$Ne as a $4^+$, but more experimental 
data are needed.

While predictions for individual nuclei in the lower $sd$ shell can be seen 
in the supplementary material \cite{SupMat}, it may be difficult to conclude definitively on 
the quality of these predictions with respect to experiment and USDB.  
Therefore we have calculated the root-mean-squared deviation from 144 
experimental levels in the $sd$-shell $Z=8-12$ isotopes.  For the shell 
model IM-SRG (USDB) interactions, we find values of 513(244)$\keV$ in 
oxygen, 446(200)$\keV$ in fluorine, 388(268)$\keV$ in neon, 
572(155)$\keV$ in sodium, and 791(106)$\keV$ in magnesium. While the 
experimental agreement for fluorine and neon is an improvement over 
the description of oxygen isotopes in Ref.~\cite{Bogn14SM}, the 
decreased accuracy for magnesium in particular is likely due to a
combination of neglected 3N forces between valence-space nucleons and 
a deterioration of the NN+3N-full Hamiltonians.

\begin{figure}[t]
\begin{center}
\minipage{0.25\textwidth}
\includegraphics[width=\linewidth,clip=]{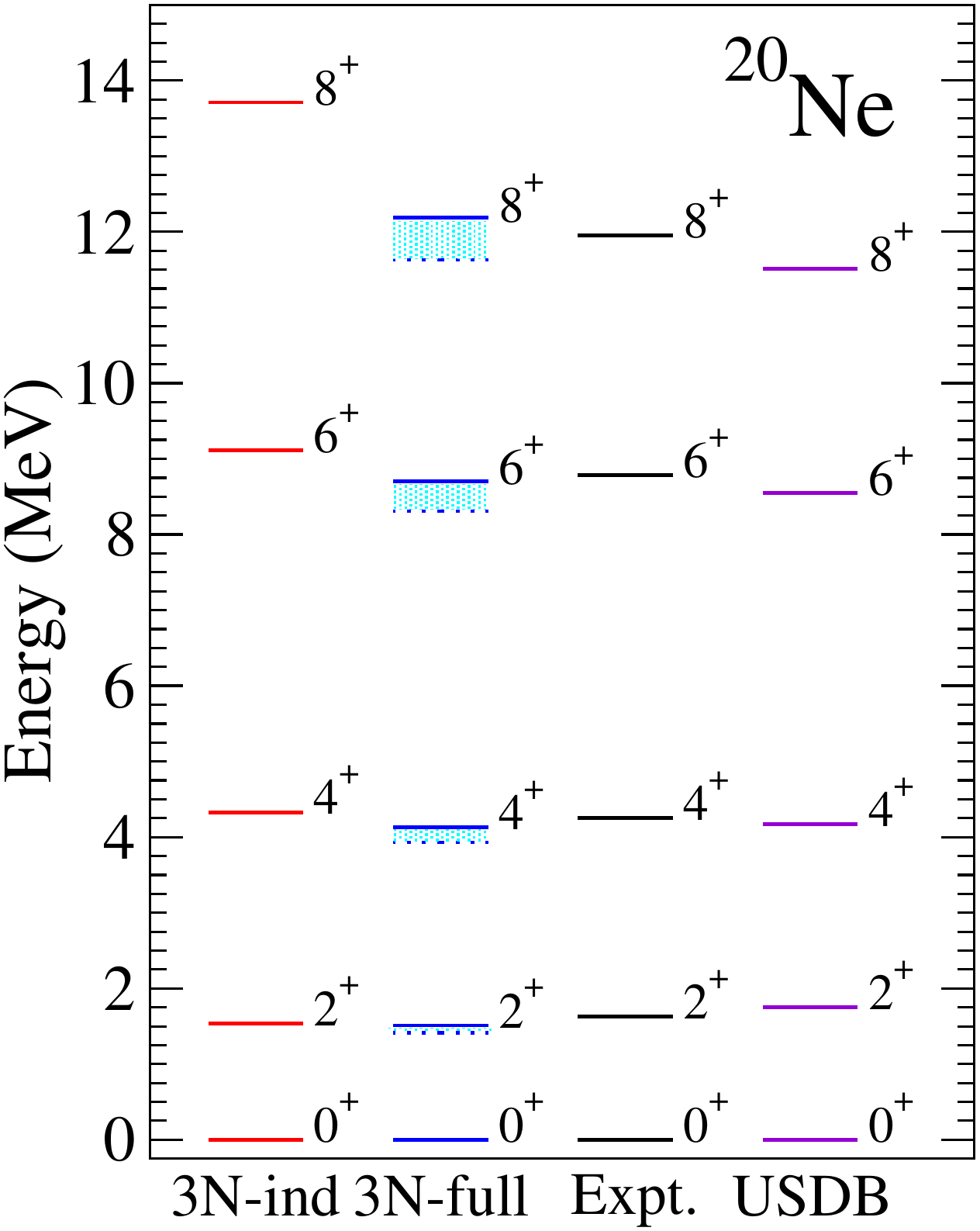}
\endminipage\hfill
\minipage{0.23\textwidth}
\includegraphics[width=\linewidth,clip=]{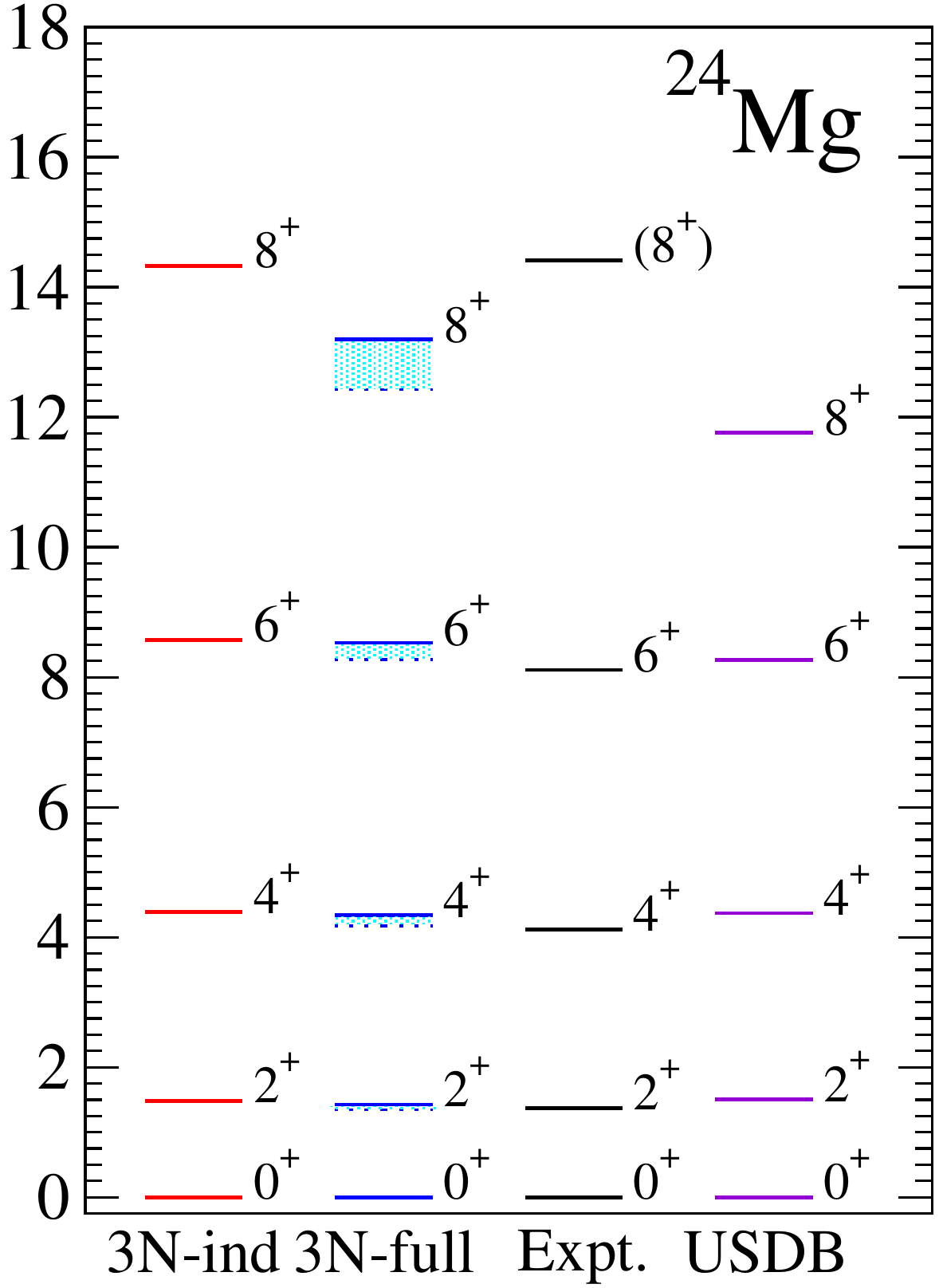}
\endminipage\hfill
\end{center}
\caption{Yrast states for deformed $^{20}$Ne and $^{24}$Mg compared 
to experimental data and phenomenological USDB predictions.}
\label{def}
\end{figure}

Finally we turn to deformation, which can be treated ab initio in light nuclei 
with Green's Function Monte Carlo \cite{Piep04A68}, with the standard or 
symplectic no-core shell model \cite{Caur01Be8,Dytr13def,Capr15def}, 
and with lattice EFT \cite{Lahd13LatEFT}, or within an EFT framework for 
heavy nuclei \cite{Pape10def,Pere15E2}. Deformation is challenging for ab 
initio methods to capture in medium-mass nuclei, where spherical 
symmetry is typically assumed, and extensions to the computationally 
demanding $m$-scheme are required for a proper treatment. Within the 
present framework, deformation can emerge naturally from valence-space 
configuration mixing, and here we investigate the extent to which this is 
realized. One key signature of deformation is the presence of a rotational 
spectrum. $^{20}$Ne and $^{24}$Mg provide classic examples of 
rotational spectra in the lower $sd$ shell, and these spectra are well 
reproduced in all calculations, as shown in Fig.~\ref{def}, though 
somewhat improved with the inclusion of 3N forces. Further evidence of 
deformation may be deduced from Fig.~\ref{gs}, where we note a 
significant discrepancy in the $^{20,22}$Ne ground-state energies 
obtained with the shell model and multi-reference calculations. This may 
be understood by considering that the multi-reference IM-SRG, which is 
built on \emph{intrinsically spherical} reference states, cannot produce a 
deformed ground state. We might expect that instead it selects the
lowest-energy state with spherical intrinsic structure, and indeed, we find 
that the energy of the first excited $0^+$ state from the valence space 
calculation aligns remarkably well with the multi-reference result in 
Fig.~\ref{gs}.

%\paragraph{Conclusions.}

In conclusion, we have presented ab initio calculations for doubly 
open-shell nuclei from $A$-dependent IM-SRG valence-space 
Hamiltonians. With initial 3N forces, excited states are in agreement with 
experiment, and with a new targeted normal ordering procedure, 
ground-state energies are improved with respect to experiment and 
large-space multi-reference IM-SRG calculations. A systematic application 
of targeted normal ordering, which better accounts for effects of 3N forces 
between valence space particles, will allow ab initio calculations 
throughout the $sd$ shell. Comparison with multi-reference IM-SRG 
indicates that the valence-space IM-SRG calculations produce deformed 
ground states in $^{20,22}$Ne and predict rotational yrast states in 
deformed $^{20}$Ne and $^{24}$Mg, illustrating that deformation can be 
captured in this ab initio framework. To further explore deformation in the 
$sd$ shell, the Magnus formulation allows straightforward evaluation of 
relevant effective valence-space operators such as quadrupole moments 
and $E2$ transitions, and ultimately extensions to  other operators will 
allow ab initio predictions for important electroweak processes such as 
neutrinoless double-beta decay 
\cite{Avig08RMP,Mene110nbb2bc,Holt13GTeff}.

\paragraph{Acknowledgments.}

We thank A.\ Calci, J.~Men\'{e}ndez, T.\ Morris, P.\ Navr\'{a}til, N.
Parzuchowski, A.~Poves, J.~Simonis, and O.~Sorlin for useful discussions 
and S.\ Binder, A.\ Calci, J.\ Langhammer, and R.\ Roth for the 
SRG-evolved NN+3N matrix elements. TRIUMF receives funding via a 
contribution through the National Research Council Canada. This work 
was supported in part by NSERC, the NUCLEI SciDAC Collaboration 
under the U.S.\ Department of Energy Grants No.~DE-SC0008533 and 
DE-SC0008511, the National Science Foundation under Grants 
No.~PHY-1404159, the European Research Council Grant No.~307986 
STRONGINT, and the BMBF under Contracts No.~06DA70471 and 
05P15RDFN1. Computations were performed with an allocation of 
computing resources at the J\"ulich Supercomputing Center, Ohio 
Supercomputer Center (OSC), and the Michigan State University High 
Performance Computing Center (HPCC)/Institute for Cyber-Enabled 
Research (iCER).

\paragraph{Note added.} Very recently Jansen et al.~\cite{Jans15def} 
applied the complementary coupled-cluster effective-interaction method to 
construct $A$-independent nonperturbative shell-model interactions also 
to explore deformation in the $sd$ shell.

\bibliography{IMSRG_FNe}

\end{document}